\documentclass{aa}
\usepackage{graphicx}
\begin{document}

\title{\bf CUrious Variables Experiment (CURVE).\\
Variable properties of the dwarf nova SS UMi.}
\author{A. Olech\inst{1}, K. Mularczyk\inst{2}
P. K\c{e}dzierski\inst{2}, K. Z{\l}oczewski\inst{2},
M. Wi{\'s}niewski\inst{1}, and K. Szaruga\inst{2}}

\offprints{A. Olech}

   \institute{Nicolaus Copernicus Astronomical Center, ul. Bartycka 18,
        00-716 Warszawa, Poland \\
        \email{(olech,mwisniew)@camk.edu.pl}
        \and
    Warsaw University Observatory, Al. Ujazdowskie 4, 00-478 Warszawa, Poland\\
        \email{(kmularcz,pkedzier,kzlocz)@astrouw.edu.pl}}

   \date{Received 3 November 2005 / Accepted ............................. 2005}

   \abstract{We report on extensive photometry of the dwarf nova SS Ursae
Minoris throughout nine months of 2004. In total, we recorded two
superoutbursts and 11 normal outbursts of the star. SS UMi has been known to
show frequent superoutbursts with a mean interval of 84.7
days. Our data suggest that the interval between successive
superoutbursts lengthened to 197 days, indicating that SS UMi entered a
period of untypical behavior manifested by a growth in the quiescent 
magnitude of the star and a series of frequent, low-amplitude, normal 
outbursts observed from July to September 2004.

The mean superhump period derived for the April 2004 superoutburst of SS
UMi is $P_{\rm sh} = 0.070149(16)$ days ($101.015\pm0.023$ min).
Combining this value with an earlier orbital period determination, we
were able to derive the period excess, which is equal to $3.5\pm1.6$\%,
and estimate the mass ratio of the binary system as equal to
$q=0.16\pm0.07$.

During the entire superoutburst, the period decreased at a rate of $\dot
P/P_{\rm sh} = -6.3(1.4) \times 10^{-5}$. However, detailed analysis of
the timings of superhump maxima seem to suggest a more complex period
change, with a decrease in the period during the first and last stages
of the superoutburst but an increase in the middle interval.

}

\titlerunning{CURVE. Variable properties of the dwarf nova SS UMi.}
\authorrunning{Olech et al.}
   \maketitle
%

\section{Introduction}

SS Ursae Minoris (PG 1551 +719) was identified as a cataclysmic variable
candidate in the Palomar Green Survey (Green et al. 1982) and, almost at
the same time was discovered, as an X-ray source E1551 +718 (Mason et al.
1982).

The quiescent spectra of the star taken by Mason et al. (1982) showed
broad and strong hydrogen and helium emission lines typical of dwarf
novae. The HeII 4686 emission line was weak and the object had no
detectable polarization (Green et al. 1982), which suggested that the
star has no strong magnetic field and thus is a typical nonmagnetic
dwarf nova. The broad emission lines have double peaked structure,
indicating high inclination and thus the possible presence of an orbital
wave in the light curve from quiescence.

The photometric behavior of SS UMi was investigated by Andronov (1986),
who observed the object photographically on 77 plates. In total, eight
possible eruptions were detected. The photographic photometry of SS UMi
was analyzed further by Richter (1989). He found that the height of the
maxima varies between 13.2 and 14.3 mag and that the star shows two
types of eruptions: long and short. The fact that faint maxima were
sometimes longer than bright ones suggested that SS UMi belongs to U Gem
class rather than SU UMa type variables.

The first CCD photometry of SS UMi was reported by Udalski (1990), who
found clear light modulations in the quiescence but no eclipses. The two
possible periods were 6.8 and 9.5 hours suggesting that one of them
might be an orbital period of the binary system and thus SS UMi is in
fact a U Gem type star. Photometric measurements of SS UMi were also
obtained by Chen et al. (1991) who observed two outbursts of the star.
Detection of the superhumps, i.e. 0.3 mag amplitude characteristic
tooth-shape light modulations during eruptions from March and September
1989 clearly showed that SS UMi belongs to the SU UMa class of dwarf
novae. Chen et al. (1991) estimated the period of the superhumps to be
around 101 minutes.

\begin{table*}[!t]
\caption{Journal of the CCD observations of SS UMi}
{\small
\begin{tabular}{||l|c|c|r|r||l|c|c|r|r||}
\hline
\hline
Date of& Start & End & Length & No. of & Date of& Start & End & Length & No. of \\
2004   & 2453000. + & 2453000. + & [hr]~ & frames & 2004   & 2453000. + & 2453000. + & [hr]~ & frames\\
\hline
Apr 13/14 & 109.54367 & 109.60431 & 1.455 & 14 &        Jul 31/01 & 218.31734 & 218.34645 & 0.699 & 7\\
Apr 14/15 & 110.52498 & 110.60747 & 1.980 & 43 &        Aug 01/02 & 219.32961 & 219.35228 & 0.544 & 12\\
Apr 15/16 & 111.44819 & 111.55250 & 2.503 & 54 &        Aug 04/05 & 222.32545 & 222.33127 & 0.140 & 4\\
Apr 18/19 & 114.56803 & 114.57752 & 0.228 & 3 &         Aug 07/08 & 225.43890 & 225.45064 & 0.282 & 5\\
Apr 19/20 & 115.44451 & 115.55515 & 2.655 & 27 &        Aug 09/10 & 227.31842 & 227.32588 & 0.179 & 6\\
Apr 20/21 & 116.46420 & 116.53175 & 1.621 & 36 &        Aug 10/11 & 228.31429 & 228.32109 & 0.163 & 4\\
Apr 21/22 & 117.43697 & 117.45411 & 0.411 & 6 &         Aug 11/12 & 229.30221 & 229.31089 & 0.208 & 5\\
Apr 22/23 & 118.47092 & 118.54493 & 1.776 & 32 &        Aug 12/13 & 230.31946 & 230.32380 & 0.104 & 3\\
Apr 23/24 & 119.38887 & 119.49275 & 2.493 & 51 &        Aug 13/14 & 231.32677 & 231.33327 & 0.156 & 4\\
Apr 25/26 & 121.31750 & 121.59130 & 6.571 & 178 &       Aug 14/15 & 232.31815 & 232.57177 & 1.356 & 27\\
Apr 26/27 & 122.32719 & 122.58720 & 6.240 & 137 &       Aug 15/16 & 233.32523 & 233.58060 & 6.129 & 101\\
Apr 27/28 & 123.34403 & 123.45042 & 2.553 & 37 &        Aug 16/17 & 234.44161 & 234.44564 & 0.097 & 2\\
Apr 28/29 & 124.32926 & 124.45396 & 2.993 & 39 &        Aug 17/18 & 235.31037 & 235.33822 & 0.668 & 11\\
Apr 29/30 & 125.38484 & 125.48146 & 2.319 & 51 &        Aug 18/19 & 236.48867 & 236.51599 & 0.656 & 13\\
Apr 30/01 & 126.37744 & 126.50035 & 2.950 & 64 &        Aug 19/20 & 237.30917 & 237.32350 & 0.344 & 10\\
May 03/04 & 129.48034 & 129.49041 & 0.242 & 4 &         Aug 21/22 & 239.31643 & 239.32572 & 0.223 & 5\\
May 05/06 & 131.48946 & 131.49493 & 0.131 & 3 &         Aug 22/23 & 240.40505 & 240.41100 & 0.143 & 4\\
May 10/11 & 136.47427 & 136.48934 & 0.362 & 6 &         Sep 01/02 & 250.31094 & 250.31941 & 0.203 & 4\\
May 11/12 & 137.42191 & 137.42374 & 0.044 & 2 &         Sep 02/03 & 251.26665 & 251.28535 & 0.449 & 10\\
May 12/13 & 138.44843 & 138.47103 & 0.542 & 7 &         Sep 03/04 & 252.27148 & 252.28902 & 0.421 & 7\\
May 14/15 & 140.41082 & 140.43849 & 0.664 & 16 &        Sep 05/06 & 254.30278 & 254.32018 & 0.418 & 5\\
May 20/21 & 146.43194 & 146.45168 & 0.474 & 10 &        Sep 06/07 & 255.26395 & 255.27509 & 0.267 & 6\\
May 23/24 & 149.43084 & 149.44656 & 0.377 & 6 &         Sep 07/08 & 256.30955 & 256.45321 & 3.448 & 40\\
May 24/25 & 150.41066 & 150.41786 & 0.173 & 5 &         Sep 08/09 & 257.29443 & 257.53091 & 5.676 & 91\\
May 25/26 & 151.44124 & 151.44550 & 0.102 & 2 &         Sep 09/10 & 258.37591 & 258.39313 & 0.413 & 7\\
Jun 03/04 & 160.38724 & 160.39526 & 0.192 & 5 &         Sep 10/11 & 259.25465 & 259.27890 & 0.582 & 11\\
Jun 30/01 & 187.35596 & 187.38250 & 0.637 & 10 &        Sep 11/12 & 260.31552 & 260.33897 & 0.563 & 11\\
Jul 02/03 & 189.44094 & 189.44500 & 0.097 & 3 &         Sep 18/19 & 267.24640 & 267.24861 & 0.053 & 3\\
Jul 03/04 & 190.47962 & 190.49031 & 0.257 & 6 &         Sep 19/20 & 268.38386 & 268.41340 & 0.709 & 11\\
Jul 04/05 & 191.38762 & 191.40979 & 0.532 & 2 &         Sep 24/25 & 273.25110 & 273.37382 & 2.945 & 50\\
Jul 06/07 & 193.36622 & 193.37928 & 0.313 & 6 &         Sep 25/26 & 274.31794 & 274.47224 & 3.703 & 64\\
Jul 08/09 & 195.33973 & 195.53558 & 4.700 & 98 &        Sep 27/28 & 276.22740 & 276.23427 & 0.165 & 4\\
Jul 09/10 & 196.33697 & 196.53760 & 0.568 & 14 &        Oct 01/02 & 280.39856 & 280.41151 & 0.311 & 6\\
Jul 11/12 & 198.34706 & 198.36742 & 0.489 & 12 &        Oct 03/04 & 282.39762 & 282.41202 & 0.346 & 7\\
Jul 13/14 & 200.35101 & 200.36211 & 0.266 & 7 &         Oct 12/13 & 291.33376 & 291.35430 & 0.493 & 10\\
Jul 15/16 & 202.46961 & 202.47820 & 0.206 & 5 &         Oct 13/14 & 292.32704 & 292.33975 & 0.305 & 6\\
Jul 17/18 & 204.34832 & 204.36139 & 0.314 & 7 &         Oct 20/21 & 299.25359 & 299.25545 & 0.044 & 2\\
Jul 18/19 & 205.34856 & 205.35433 & 0.138 & 3 &         Nov 06/07 & 316.22524 & 316.22569 & 0.011 & 1\\
Jul 19/20 & 206.39249 & 206.40227 & 0.235 & 4 &         Nov 14/15 & 324.17099 & 324.26325 & 2.214 & 50\\
Jul 22/23 & 209.36131 & 209.39549 & 0.820 & 10 &        Nov 20/21 & 330.19057 & 330.21009 & 0.468 & 9\\
Jul 23/24 & 210.34668 & 210.35964 & 0.311 & 7 &         Nov 21/22 & 331.68761 & 331.69816 & 0.253 & 6\\
Jul 28/29 & 215.32681 & 215.54904 & 5.333 & 148 &       Dec 01/02 & 341.21800 & 341.23103 & 0.313 & 6\\
Jul 29/30 & 216.33435 & 216.56104 & 5.441 & 89 &        Dec 06/07 & 346.26354 & 346.27634 & 0.307 & 6\\
Jul 30/31 & 217.33269 & 217.55530 & 5.343 & 92 &        Dec 08/09 & 348.24499 & 348.25133 & 0.152 & 4\\
\hline
\hline
\end{tabular}}
\end{table*}

Time-resolved specroscopy of SS UMi and six other SU UMa stars was
obtained by Thorstensen et al. (1996). The variations in the positions
of H$\alpha$ emission lines allowed determination of the orbital period of
the binary, which is equal to $P_{\rm orb} = 0.06778$ days ($97.6\pm1.5$
min).

Another superoutburst of SS UMi was observed by Kato et al. (1998). They
observed the star on four nights of April 1998 eruption. During two
nights the star showed clear superhumps with a mean period of $P_{\rm
sh}=0.0699\pm0.0003$ days ($100.7 \pm 0.4$ min).

Kato et al. (2000) analyzed 375 visual and CCD observations of SS UMi
taken from February 18, 1999 to June 17, 2000 and reported by VSNET
observers. Inspection of the global light curve allowed them to find
that the intervals between successive superoutbursts are in the range of
82 -- 86 days with a mean value of 84.7 days. They also noticed that
there are five normal outbursts bewteen two successive superoutbursts,
meaning that the mean cycle length is around 11 days.

\section{ER UMa stars}

The first members of the ER UMa class objects were discovered in the mid
1990s (ER UMa, RZ LMi, V1159 Ori, and DI UMa). These are systems
characterized by an extremely short supercycle (20-60 days), a short
interval between normal outbursts (3-4 days), and small amplitude
($\sim3$ mag) of superoutbursts (Kato and Kunjaya 1995, Robertson et al.
1995, Patterson et al. 1995). Subsequently, one more object with similar
characteristics was discovered (IX Dra - Ishioka et al. 2001, Olech et
al. 2004a).

These objects seemed to be very unusual compared to normal SU UMa stars.
Ten years ago, the ordinary SU UMa star with the shortest supercycle of
134 days was YZ Cnc (Patterson 1979, Shafter and Hessman 1988). Thus
supercycles of ER UMa stars were about 3-4 times shorter. However, when
describing the results of the CBA observational campaign for V1159 Ori,
Patterson et al. (1995) claimed that there is no reason to introduce
a new class of variable stars. Simply, the observable traits of ER
UMa-type stars seem to be consistent with garden-variety SU UMa stars.
They follow the Kukarkin-Parengo relation connecting the amplitude of
the outburst with recurrence time between normal outbursts and the
Bailey relation connecting decay times from the normal eruptions and orbital
period of the binary. They simply appear to be normal SU UMa stars with
greater activity and greater luminosity due to their higher-mass
transfer rates (Osaki 1996).

However, careful inspection of the diagram with recurrence intervals for
supermaxima vs. normal maxima showed a significant gap between normal SU
UMa stars and ER UMa-type variables (see Fig. 18 of Paterson et al.
1995).

As we can see, SS UMi with a supercycle of 85 days lies in the
transition area between extremely active dwarf novae and normal SU UMa
stars. Investigating the behavior and characteristics of this star might
help us to better understand this class of objects. This was the reason
for including SS UMi into the primary targets observed within the CURVE
(Olech et al. 2003a,b). In this work we report the results of a nine
month observational campaign performed in 2004.

\section{Observations and data reduction}

Observations of SS UMi reported in the present paper were obtained
during 88 nights between April 13, 2004 and December 8, 2004 at the
Ostrowik station of the Warsaw University Observatory. The data was
collected using the 60-cm Cassegrain telescope equipped with a
Tektronics TK512CB back-illuminated CCD camera. The scale of the camera
was 0.76"/pixel providing a $6.5'\times 6.5'$ field of view. The full
description of the telescope and camera was given by Udalski and Pych
(1992). We monitored the star in ``white light''.  We did not use any
filters in order to shorten the exposures to minimize guiding errors,
due to the temporary lack of an autoguiding system, after a recent
telescope renovation. The exposure times were from 100 to 150 seconds
during the bright state and from 180 to 300 seconds at minimum light.

A full journal of our CCD observations of SS UMi is given in Table 1. In
total, we monitored the star during 105.374 hours and obtained 2021
exposures. All the data reductions were performed using a standard
procedure based on the IRAF\footnote{IRAF is distributed by the National
Optical Astronomy Observatory, which is operated by the Association of
Universities for Research in Astronomy, Inc., under a cooperative
agreement with the National Science Foundation.} package and the profile
photometry was derived using the DAOphotII package (Stetson 1987).

Relative, unfiltered magnitudes of SS UMi were determined as the
difference between the magnitude of the variable and the intensity
averaged magnitude of the two comparison stars: C1 (${\rm RA} =
15^h51^m59.5^s$, Decl.$= +71^\circ 44'25.6"$, $V=12.553$, $B-V=0.883$)
and C2 (${\rm RA} = 15^h52^m07.3^s$, Decl.$= +71^\circ 46'34.6"$, 
$V=13.226$, $B-V=1.052$). The comparison stars and the variable are
marked in the chart displayed in Fig. 1. The basic properties of
comparison stars are taken from the catalog of Henden and Honeycutt
(1997).

   \begin{figure}[!h]
   \centering
\includegraphics[scale=.49]{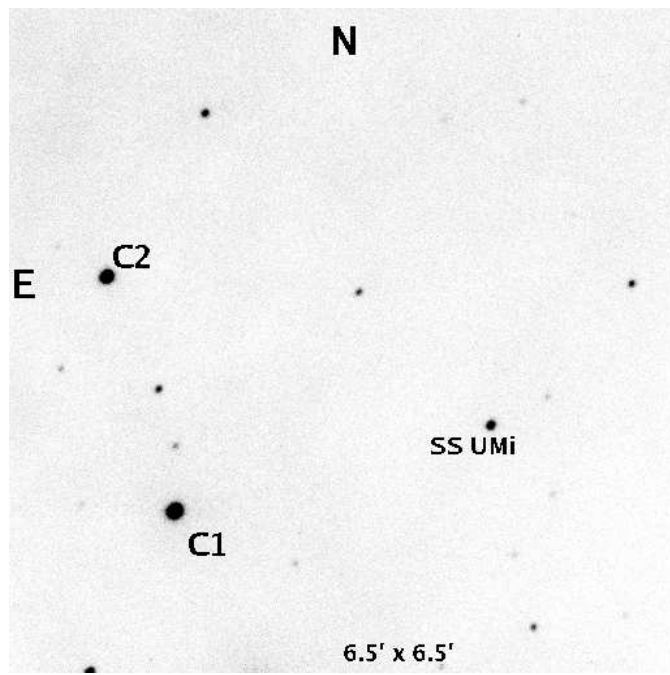}
      \caption{Finding chart for SS UMi covering a region of $6.5 \times
6.5$ arcminutes. The positions of the comparison stars are shown. North is
up, East is left.
              }
         \label{Fig1}
   \end{figure}

   \begin{figure*}[!t]
   \centering
\includegraphics[bb=90 375 520 620, scale=0.96]{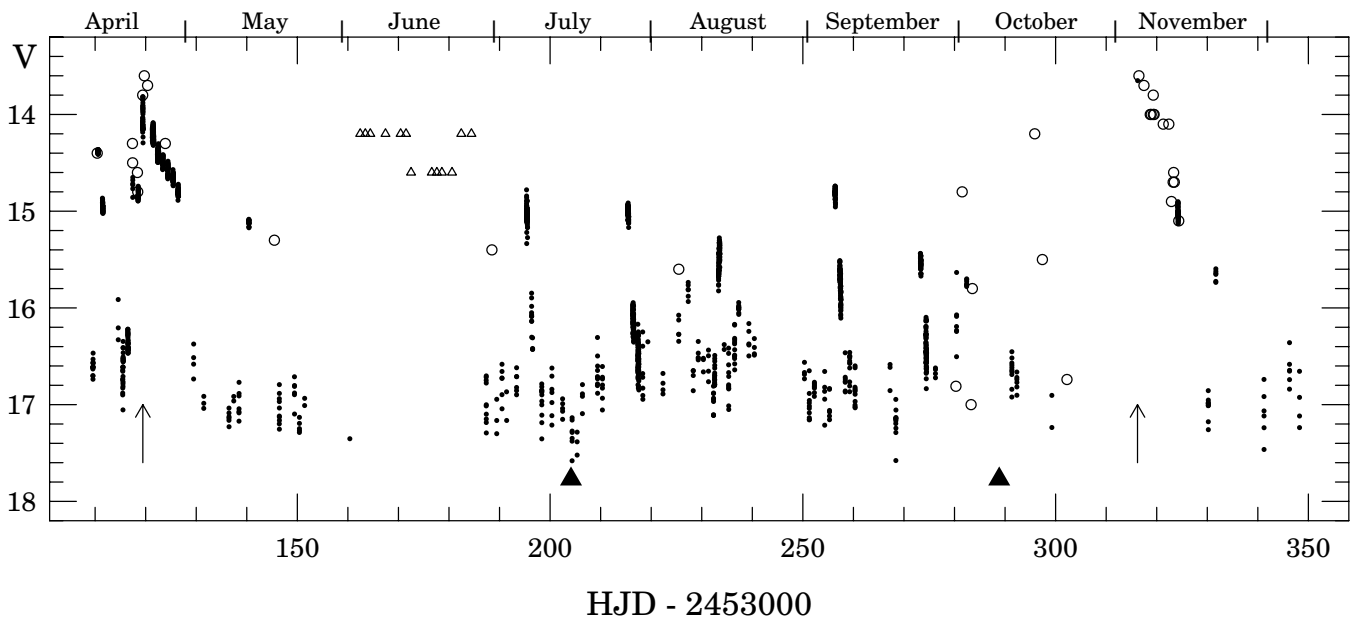}
      \caption{The general photometric behavior of SS UMi during
the whole 2004 campaign. Visual estimates collected by AAVSO observers
are shown as open circles and the CCD observations described in this
paper as black dots. The open triangles correspond to the 'fainter than'
estimates from AAVSO. Big filled triangles point to the expected positions
of superoutbursts assuming 84.7 days supercycle length. Two arrows show the
moments of maximum magnitude of the recorded superoutbursts.}
         \label{Fig2}
   \end{figure*}

The typical accuracy of our measurements varied between 0.006 and 0.024
mag in the bright state and between 0.009 and 0.2 mag in the minimum
light. The median value of the photometry errors was 0.009 and 0.045
mag, respectively.

\section{Global light curve}

Almost all our observations were made in "white light" in order to
obtain precise photometry of the star in quiescence. However, on the
night of April 22/23, 2004 we also made $V$, $R$, and $I$ photometry of
the variable and comparison stars. Using the magnitude data from Henden
and Honeycutt (1995), we were able to transform our instrumental
magnitudes to the standard $V$ system.

Such a transformed global light curve of SS UMi spanning eight months of
observations is shown in Fig, which shows both our own CCD observations
and those in the visual made by American Association of Variable Stars
Observers (AAVSO), including their CCD magnitude estimates and also the
upper magnitude limits in case of non detection of the variable. It is
clear that the brightness of SS UMi varies from around 17.2 mag at
quiescence to around 13.7 mag at the maximum of superoutburst. During
the brightest normal outbursts the star can be found at a magnitude of
around 14.4.

The vertical arrows point to the maximum brightness of the two
superoutbursts clearly visible in the light curve. They are separated by
197 days, which is much longer than the mean supercycle length of 84.7
days determined by Kato et al. (2000).

As we can see, the lack of our CCD observations during June nights, HJD
from 2453161 to 2453187, may suggest that another superoutburst occurred
in that time. This is, however, unlikely for two reasons. First, it
would suggest that the supercycle of SS UMi shortened to about 50 days,
and second, the observations of AAVSO clearly indicate that during June
nights the star was fainter than 14.2 -- 14.6 mag and thus was able to
show only ordinary outbursts but no superoutburst. The big filled
triangles in Fig. 2 denote the expected moments of maxium brightness of
the missing superoutbursts assuming a 84.7-day supercycle of SS UMi.

\section{April 2004 superoutburst}

The April 2004 superoutburst started on Apr. 20/21 with a normal
outburst, the so-called precursor. After an initial rise in brightness,
the magnitude of the star decreased slightly and then the tidal
instability triggered the superoutburst manifested by the appearance
of the superhumps and an increase in the brightness. Starting with the
night of Apr. 23/24, SS UMi entered the plateau phase. During this
period clear superhumps were present and the brightness of the star was
decreasing at a rate of 0.12 mag/day. Around May 1st, the magnitude
started to decrease much faster, indicating termination of the
superoutburst. On May 4/5 the star was again at its quiescent magnitude.

\subsection{Superhumps}

The light curves from individual nights of the April 2004 superoutburst
of SS UMi are shown in Fig. 3. Clear superhumps were present in the
light curve of the star from Apr. 22/23 to Apr. 30/01. During the first
night, before reaching maximum brightness, the amplitude of the
superhumps was only 0.13 mag. A night later, the superhumps were fully
developed with their characteristic tooth-shape and an amplitude of 0.33
mag. On Apr. 24/25 the shape of the superhumps did not change, but the
amplitude decreased to 0.15-0.20 mag depending on the hump. The next
night the amplitude stayed at the same level, but clear secondary humps
became visible. In the interval Apr. 26/27 -- Apr. 28/29, the amplitude
of the light variations was around 0.15 mag and secondary humps became
more pronounced. On Apr. 30/01 the superhumps had a smaller amplitude
than 0.1 mag and after that night the plateau phase ended.

   \begin{figure}[!t]
   \centering
\includegraphics[bb=150 260 520 710,scale=.95]{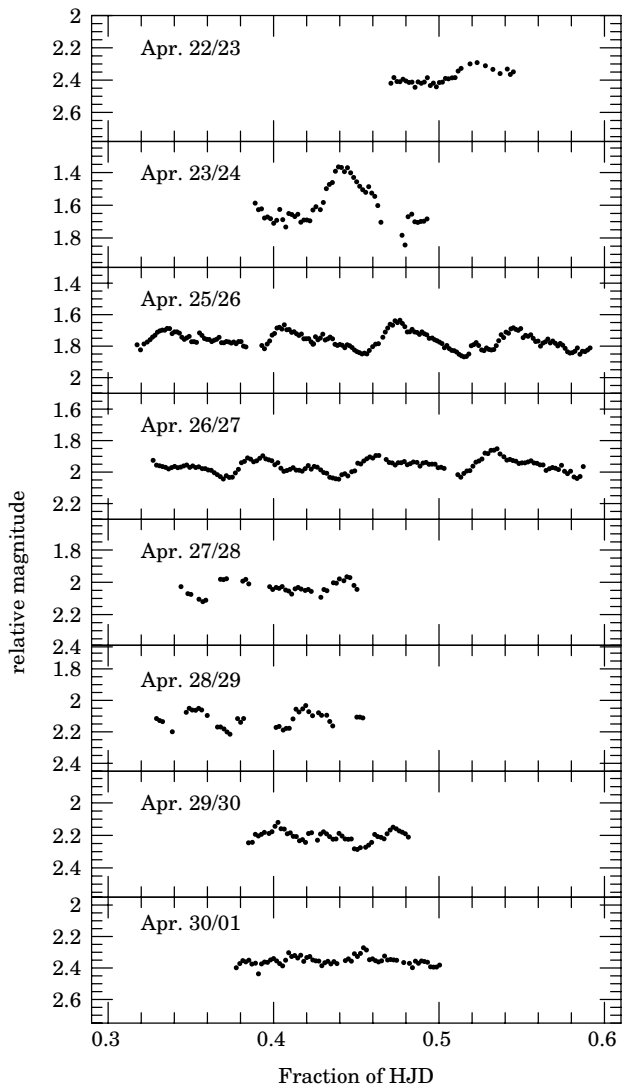}
      \caption{The light curves of SS UMi during its April 2004 superoutburst.
              }
         \label{Fig3}
   \end{figure}

   \begin{figure}
   \centering
\includegraphics[bb=150 390 520 610,scale=.8]{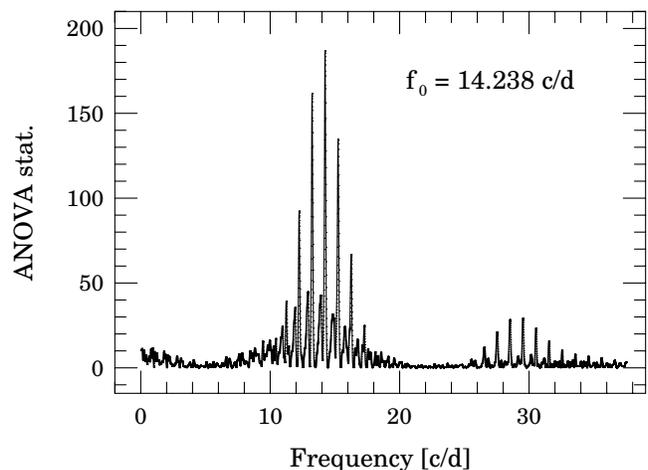}
      \caption{The ANOVA power spectrum of the light curve of SS UMi from
its 2004 April superoutburst.
              }
         \label{Fig4}
   \end{figure}

From each light curve of SS UMi in superoutburst, we removed the first or
second order polynomial and analyzed them using {\sc anova} statistics
with two harmonic Fourier series (Schwarzenberg-Czerny 1996). The
resulting periodogram is shown in Fig. 4. The most prominent peak is
found at a frequency of $f=14.238\pm0.020$ c/d, which corresponds to the
period of $P_{sh}=0.07023(10)$ days ($101.13\pm0.14$ min).  The harmonic
peak at 28.5 c/d is real and leads to the presence of secondary
humps in the light curve.

From each light curve of the superoutburst we removed variablility 
corresponding to the superhump period. Such a prewhitened light curve
was again analyzed by {\sc anova} statistics. No significant peak was
recoreded in the computed power spectrum.

\subsection{The $O - C$ analysis}

To check the stability of the superhump period and to determine its
value, we constructed an $O-C$ diagram. We decided to use the timings of
primary maxima, because they were almost always high and clearly
detectable in the light curve of the variable. In the end, we were able
to determine 17 moments of maxima, which are listed in Table 2
together with their errors, cycle numbers $E$, and $O-C$ values.

\begin{table}[!h]
\caption{Timings of maxima in the light curve of SS UMi during its 2004
April superoutburst.}
\begin{center}
\begin{tabular}{||c|c|r|r||}
\hline
\hline
Cycle $E$ & HDJ - 2453000 & Error~ & $O-C$~ \\
\hline
\hline
0 & 118.5230 & 0.0020 & $-0.0912$\\
13 & 119.4418 & 0.0025 & $+0.0070$ \\
40 & 121.3358 & 0.0030 & $+0.0074$\\
41 & 121.4054 & 0.0035 & $-0.0004$\\
42 & 121.4745 & 0.0022 & $-0.0153$\\
43 & 121.5451 & 0.0025 & $-0.0088$\\
55 & 122.3935 & 0.0020 & $+0.0857$\\
56 & 122.4619 & 0.0025 & $+0.0608$\\
57 & 122.5326 & 0.0020 & $+0.0687$\\
69 & 123.3720 & 0.0035 & $+0.0350$\\
70 & 123.4432 & 0.0030 & $+0.0500$\\
83 & 124.3518 & 0.0025 & $+0.0028$\\
84 & 124.4198 & 0.0023 & $-0.0278$\\
98 & 125.4026 & 0.0015 & $-0.0172$\\
99 & 125.4723 & 0.0018 & $-0.0236$\\
112 & 126.3829 & 0.0030 & $-0.0422$\\
113 & 126.4542 & 0.0020 & $-0.0258$\\
\hline
\hline
\end{tabular}
\end{center}
\end{table}

The least-square linear fit to the data from Table 2 gives the
following ephemeris for the maxima:

\begin{equation}
{\rm HJD}_{\rm max} =  2453118.5294(12) + 0.070147(17) \times E
\end{equation}

The $O-C$ values computed according to the ephemeris (1) are listed in
Table 2 and also shown in Fig. 5. It is clear that the superhump period
shows a slightly decreasing trend, confirmed by the second-order
polynomial fit to the moments of maxima, which gives the following
ephemeris (shown as a solid line in Fig. 5):

\begin{eqnarray}
{\rm HJD}_{\rm max} = 2453118.5240(17) + 0.070413(60) \times E & \\
- 2.22(0.48) \cdot 10^{-6} \times E^2 & \nonumber
\end{eqnarray}

Finally, we conclude that the period of the superhumps was not stable
during the April 2004 superoutburst of SS UMi; and in the interval
covered by our observations it is roughly described by a decreasing
trend with a rate of $\dot P/P_{\rm sh} = -6.3(1.4) \times 10^{-5}$. The
combination of both our superhump period determinations returned its
mean value to $P_{\rm sh} = 0.070149(16)$ days
($101.015\pm0.023$ min).

   \begin{figure}
   \centering
\includegraphics[bb=155 270 520 510,scale=.82]{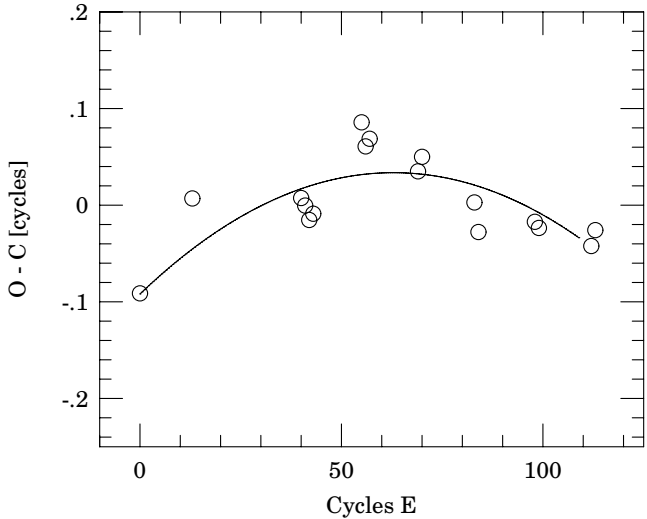}
      \caption{The $O-C$ diagram for the superhump maxima of SS UMi
during its April 2004 superoutburst. The solid line corresponds to the fit
resulting from the ephemeris (2).
              }
         \label{Fig5}
   \end{figure}

\section{Discussion}

\subsection{Basic parameters of the binary system}

Knowing the exact value of the superhump period and orbital period of
the binary system determined by Thorstensen et al. (1996), we can
compute the period excess defined as $\Delta \epsilon = (P_{\rm sh} -
P_{\rm orb})/P_{\rm orb}$. In the case of SS UMi it is equal to
$3.5\pm1.6$\%. The relatively large error comes mainly from the 1.5 min
error in the Thorsensen et al. (1996) determination of the orbital
period. The obtained value is typical for the SU UMa stars with orbital
periods of around 98 minutes, indicating that SS UMi follows the Stolz
and Schoembs (1984) relation between the period excess and orbital
period of the binary.

Superhumps occur at a period slightly longer than the orbital period of
the binary system. They are most probably the result of accretion disc
precession caused by the gravitational perturbations from the secondary.
These perturbations are most effective when the disc particles moving on
eccentric orbits enter the 3:1 resonance. Then the superhump period is
simply the beat period between orbital and precession rate periods
(Osaki 1996).

Assuming the known disk size at 3:1 resonance, it is easy to correlate
the mass ratio of the system $q=M_{\rm secondary}/M_{\rm WD}$ with the
period excess $\epsilon$:

\begin{equation}
\epsilon\approx\frac{0.23q}{1+0.27q}
\end{equation}

\noindent This relation allows estimation of the mass ratio $q$ in the
dwarf nova system based only on observational determination of the
orbital and superhump periods. In the case of SS UMi, this mass ratio is
equal to $0.16\pm0.07$.

\subsection{Superhump period change}

Until the mid 1990's, all members of the SU UMa group seemed to show
only negative superhump period derivatives (Warner 1995). This was
interpreted as a result of disk shrinkage during the superoutburst, thus
lengthening its precession rate (Lubow 1992). This picture became more
complicated when the first stars with $\dot P>0$ were discovered.
Positive period derivatives were observed only in stars with short
superhump periods close to the minimum orbital period for hydrogen rich
secondary (e.g. SW UMa - Semeniuk et al. 1997, WX Cet - Kato et al.
2001a, HV Vir - Kato et. al 2001b) or for stars below this boundary
(e.g. V485 Cen - Olech 1997, 1RXS J232953.9+062814 - Uemura et al.
2002).

Two years ago, Olech et al (2003a) investigated the $O-C$ diagrams for KS
UMa, ER UMa, V1159 Ori, CY UMa, V1028 Cyg, RZ Sge, and SX LMi and claimed
that most (probably almost all) SU UMa stars show decreasing superhump
periods at the beginning and the end of superoutburst but increasing
period in the middle phase. This hypothesis was quickly confirmed by
observations of the June 2004 superoutburst of TT Boo, which showed the
following period derivatives $(-52.3\pm1.3) \times 10^{-5}$,
$(12.3\pm4.8) \times 10^{-5}$ and $(-6.2\pm0.9) \times 10^{-5}$, from
the beginning to end of the superoutburst, respectively (Olech et al.
2004b).

Careful inspection of Fig. 5 suggests that in the case of SS UMi, we
encounter the same situation as in TT Boo. The $O-C$ residuals may be
fitted with simple parabola, as in Fig. 5, but it is clear that
this fit is only a rough estimation of global behavior. In fact, it is
safer to say that in the case of the April 2004 superoutburst of SS UMi,
the timings of superhump maxima seem to suggest more complex period
change, with a decrease in the period during the first and last stages
of the superoutburst, but an increase in the middle interval. The fitting
of parabolas to the to the cycle intervals 0--43, 13--57, and
40--113 gives the period derivatives of $(-54\pm21) \times 10^{-5}$,
$(+27\pm12) \times 10^{-5}$, and $(-7.5\pm3.7) \times 10^{-5}$,
respectively. As we can see, these values are quite similar to the
preiod derivatives determined for corresponding phases of the superoutburst
of TT Boo (Olech et al. 2004b).

Very recently, Uemura et al. (2005) have suggested that superhump period
change might be connected with a presence of the precursor in the light
curve of the superoutburst. In the case of a superoutburst without the
precursor, superhump period derivatives tend to be larger than those in
precursor-type eruptions. The precursor-type April 2004 superoutburst of
SS UMi, characterized by large period changes similar to these observed
in TT Boo, seems not to fit the scenario proposed by Uemura et al.
(2005).

\subsection{Supercycle length and mass transfer}

In normal SU UMa stars, the dependence of the supercycle length $t_S$
on the mass transfer rate $\dot M$ is an U-shaped curve with a broad
minimum at 80--85 days (Osaki 1995a,b). In fact, the supercycle
length $t_S$ consists of two parts: $t_{d}$, which is the duration of
the superoutburst, and $t_{\rm wait}$, which corresponds to the time
between the end of the superoutburst and begining of the succesive
superoutburst. In ordinary SU UMa stars, $t_{\rm wait}$ is typically around
one hundred days and $t_d$ is between 10 and 15 days.

With increasing mass transfer ($\dot M > 3 \cdot 10^{16}~{\rm
g\cdot s}^{-1}$ but still below the critical value above which the star
becomes a permanent superhumper), the supercycle starts to lengthen again.
However, this time it is not due to the long $t_{\rm wait}$ but to the
long duration of the superoutburst $t_d$. This is caused by the
quasi-steady state of the accretion disk, because mass-transfer rate
$\dot M$ approaches very near to the mass accretion rate from the disk
to the primary.

The standard thermal-tidal instability model is unable to produce stars
with supercycles shorter than 80 days. Thus for explaining the behavior
of ER UMa stars, with supercycles between 20 and 60 days, Osaki
(1995a,b) assumed that the tidal torques are weaker in such systems.
This results in a shorter duration of the superoutburst and larger disk
radius at its termination as an effect of less angular momentum removed
during such a short superoutburst. The angular momentum reservoir
of the disk could then be refilled in a shorter time and another
superoutburst might start more quickly.

   \begin{figure}
   \centering
\includegraphics[bb=135 365 520 660,scale=.78]{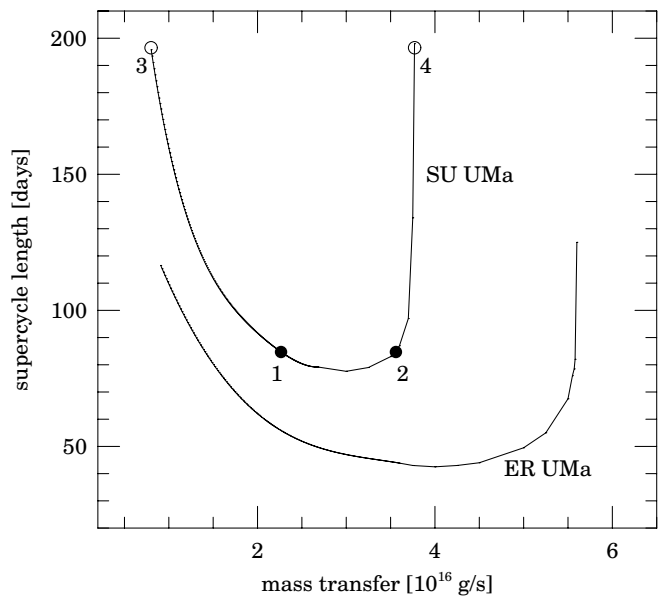}
      \caption{The dependences between supercycle length and mass
transfer in normal SU UMa stars and ER UMa variables.
              }
         \label{Fig6}
   \end{figure}

Figure 6 shows the U-shaped dependence between $t_S$ and $\dot M$ for
normal SU UMa dwarf novae and ER UMa stars taken from Osaki (1995a,b).
The filled circles denoted with numbers '1' and '2' correspond to the
two possible positions of SS UMi for a supercycle of 84.7 days. The open
circles marked by numbers '3' and '4' denote the possible positions of
SS UMi after lengthening of the supercycle to 197 days.

The possible scenarios for SS UMi correspond to the  transitions $1
\rightarrow 4$, $2 \rightarrow 4$, $1 \rightarrow 3$ and $2 \rightarrow
3$. The first two are consistent with a mass-transfer rate increasing
very close to the critical limit above which a star becomes a permanent
superhump object. In this case, the lengthening of the supercycle length
should be caused by the long duration of the superoutbursts. This contradicts
our observations, which show that both April and November
superoutbursts lasted about 15 days.

On the other hand, the transitions $1 \rightarrow 3$ and $1 \rightarrow
2$ are consistent with $t_d$ and $t_{\rm wait}$ times indicated by the
global light curve of SS UMi shown in Fig. 2. But in this case, we
should assume the decrease in the mass transfer rate by a factor of 2-4.
It is not justified by the behavior of the star between superoutbursts,
when we observed an increase in the quiescent magnitude by about
0.2--0.3 mag and increased frequency of normal eruptions, which was most
probably the reaction of the system to enhanced mass flow to the
accretion disk. The question why the star expelled the matter by a series
of frequent and low-amplitude normal outbursts observed from July to
September 2004, and  not by one or two superoutbursts, remains open.

\subsection{Another possibility for SS UMi?}

Another question arising in the case of SS UMi concerns its typical
state. It might be that SS UMi is a normal SU UMa star with a supercycle
length of 197 days and behavior observed in 1999-2000, with a supercycle
of 85 days, was atypical.

According to Kato et al. (2000), in 1999-2000 SS UMi had a supercycle of
84.7 days, a cycle of 11.0 days, and a mean amplitude of normal outbursts
of 2.7 mag. In 2004, it switched to a supercycle of 197 days. Due to the
gaps in the observational coverage, we can not estimate the
normal cycle length precisely. But between the superoutbursts from April and
November, we recorded at least 11 normal outbursts with a mean amplitude
of 2.0 mag. Taking the length of the gaps into account, we estimate that
the number of normal eruptions could reach the level of 15, indicating
a cycle length of around 12-13 days.

The amplitudes of the dwarf novae follow the famous Kukarkin-Parenago
relation (Kukarkin and Parenago 1934, Warner 1987) in the form:

\begin{equation}
A_n {\rm (mag)} = 0.70 + 1.90\log T_n {\rm (d)}
\end{equation}

\noindent where $A_n$ is an amplitude of normal outburst and $T_n$ is a
mean interval between two successive eruptions. Figure 7 shows this
relation as a solid line. It is clear that only the 1999-2000 behavior
fits the Kukarkin-Parenago relation well.

Another empirical relation followed by dwarf novae connects
their cycle and supercycle lenghts (Patterson et al. 1995, Warner 1995,
Olech et al. 2004a):

\begin{equation}
\log T_s = 1.31 + 0.644 \log T_n
\end{equation}

   \begin{figure}
   \centering
\includegraphics[bb=135 250 520 510,scale=.82]{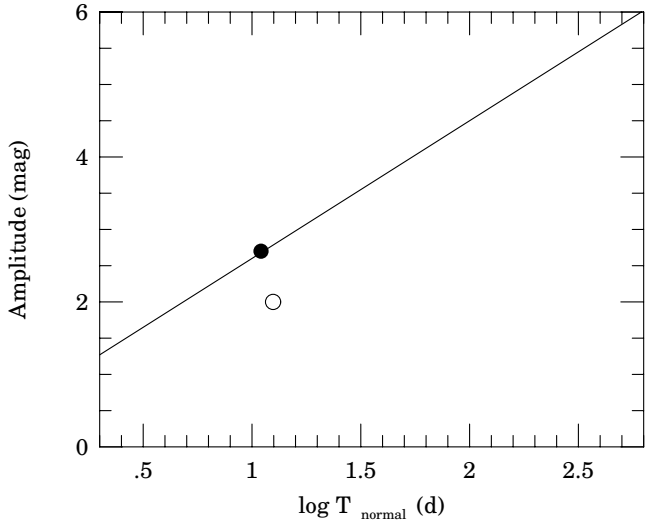}
      \caption{The Kukarkin-Parenago relation between amplitude of
the outburst and cycle length for dwarf novae. The filled and open
circles denote the positions of SS UMi for 1999-2000 and 2004,
respectively.
              }
         \label{Fig7}
   \end{figure}

   \begin{figure}
   \centering
\includegraphics[bb=135 250 520 510,scale=.82]{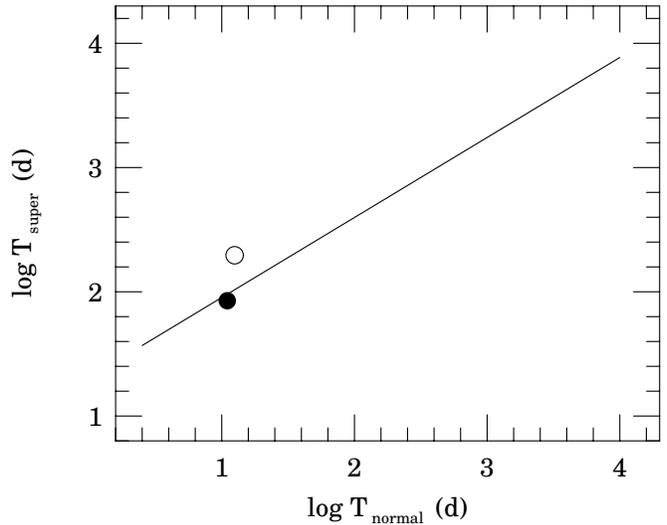}
      \caption{The relation between amplitude supercycle and cycle lengths 
for dwarf novae. The filled and open
circles denote the positions of SS UMi for 1999-2000 and 2004,
respectively.
              }
         \label{Fig8}
   \end{figure}

\noindent Figure 8 shows this relation  and again  the state from
1999-2000 seems to be more typical for dwarf novae than behavior from
2004.

\section{Summary}

The nine-month observational campaign of SS UMi performed in 2004 
allowed us to draw the following conclusions:

\begin{enumerate}

\item The supercycle of SS UMi lengthened from 84.7 days in 1999--2000
to 197 days in 2004, indicating that SS UMi entered a period of atypical
behavior manifested by a growth in the quiescent magnitude of the star
and series of frequent, low amplitude normal outbursts observed from
July to September 2004.

\item Detailed photometry during the April 2004 superoutburst of SS UMi
allowed us to determine its superhump period to be equal to $P_{\rm sh}
= 0.070149(16)$ days.

\item The combination of the superhump and orbital period determinations
allowed us to derive the period excess, which is equal to
$\epsilon=3.5\pm1.6$\%, and estimate the mass ratio of the binary system
as equal to $q=0.16\pm0.07$.

\item The superhump period was not constant during the April 2004
superoutburst of SS UMi, and its behavior is consistent with the scenario
of decreasing superhump period in the first and last stages of
superoutburst but increasing in the middle.

\item All observational properties of SS UMi can be explained using the
standard thermal-tidal instability model without assuming weaker tidal
torques as is necessary in describing the behavior of ER UMa stars.

\end{enumerate}

\begin{acknowledgements} 

We would like to thank the referee Prof. Yoji Osaki for valuable remarks
and pointing out the mistake in our interpretation. We are also grateful
to Prof. J\'ozef Smak and Dr. Grzegorz Stachowski for reading and
commenting on the manuscript. We gratefully acknowledge the generous
allocation of time at the Warsaw Observatory 0.6-m telescope. This work
was partially supported by KBN grant number 1~P03D~006~27 to A. Olech
and used the on-line service of the AAVSO. 

\end{acknowledgements}

\end{document}